\documentclass[showpacs,twocolumn]{revtex4}
\usepackage{graphicx}

\begin{document}
\title{Wormhole dynamics in spherical symmetry}
\author{Sean A. Hayward}
\affiliation{Center for Astrophysics, Shanghai Normal University, 100 Guilin
Road, Shanghai 200234, China}
\date{30th March 2009}

\begin{abstract}
A dynamical theory of traversable wormholes is detailed in spherical symmetry. 
Generically a wormhole consists of a tunnel of trapped surfaces between two 
mouths, defined as temporal outer trapping horizons with opposite senses, in 
mutual causal contact. In static cases, the mouths coincide as the throat of a 
Morris-Thorne wormhole, with surface gravity providing an invariant measure of 
the radial curvature or ``flaring-out''. The null energy condition must be 
violated at a wormhole mouth. Zeroth, first and second laws are derived for the 
mouths, as for black holes. Dynamic processes involving wormholes are reviewed, 
including enlargement or reduction, and interconversion with black holes. A new 
area of wormhole thermodynamics is suggested. 
\end{abstract}
\maketitle

\section{Introduction}
Space-time wormholes, short cuts between otherwise distant or unconnected 
regions of the universe, have become a popular research topic since the 
influential paper of Morris \& Thorne \cite{MT}. Early work was reviewed in the 
book of Visser \cite{Vis} and there is an extensive recent review by Lobo 
\cite{Lob}. The Morris-Thorne study was restricted to static, spherically 
symmetric space-times, and initially there were various ad hoc attempts to 
generalize the definition of wormhole by inserting time-dependent factors into 
the metric, despite the problem that such metrics become singular precisely at 
the throat, so do not necessarily describe a traversable wormhole in any sense. 

A more geometrically founded generalization was proposed by the author 
\cite{wh1} in terms of trapping horizons, also associated with black holes. 
This allows a substantial body of theory developed for black holes to be 
applied to wormholes. That reference deferred details to a longer article, of 
which this is a version, ten years late. It seems timely if only due to a 
recent resurgence of ad hoc approaches. 

A Morris-Thorne wormhole throat, at a given static time, is a minimal surface 
in the static hypersurface, i.e.\ locally minimizing area among surfaces in the 
hypersurface. Their ``flaring-out'' condition expresses strict minimality 
\cite{Vis}. A natural generalization is a spatial surface which is minimal in 
some spatial hypersurface. It is easy to show that, except in the doubly 
marginal case, this is a (future or past) trapped surface, more usually 
associated with black or white holes. Since this is a generic condition, a 
generic wormhole must consist of a space-time region. Then it is natural to 
look for the boundaries of this region, which one would expect to be trapping 
horizons, i.e.\ composed of marginal surfaces, which are extremal in null 
hypersurfaces. For a two-way traversable wormhole, there should be two temporal 
boundaries in mutual causal contact. Prosaically, the wormhole consists of a 
{\em tunnel between two mouths}. In static cases, the tunnel shrinks away and 
the two mouths coincide as the throat. In non-static cases, ``throat'' 
evidently means different things to different people, so the terminology will 
be avoided here. 

This viewpoint has various consequences. Firstly, a Morris-Thorne wormhole 
throat is a double trapping horizon, which will generally bifurcate under a 
dynamic perturbation, such as someone crossing it. This raises the issues of 
stability and, if stable, maintenance, i.e.\ returning a perturbed wormhole to 
a static state. Also, since black holes may also be defined locally in terms of 
trapping horizons, it is possible for a wormhole to collapse to a black hole, 
or for a black hole to be converted to a traversable wormhole. Concrete 
examples of such processes, both analytical and numerical, in toy models and 
full Einstein gravity, were given in a series of papers 
\cite{wh2,wh3,wh4,wh5,wh6,wh7,wh8,wh9,wh10}. 

For pedagogical reasons, this article will be restricted to spherical symmetry, 
though everything can be generalized as outlined in the original reference 
\cite{wh1}. Einstein gravity will be assumed, though the key ideas can be 
generalized to other metric-based theories and other dimensions. Section II 
reviews the necessary geometrical ideas, section III defines wormhole mouths, 
section IV checks the static limit, and section V derives some basic laws of 
wormhole dynamics and cites examples. 

\section{Geometry}
In spherical symmetry, the area $A$ of the spheres of symmetry is a geometrical 
invariant. It is convenient to use the area radius $r=\sqrt{A/4\pi}$, so that
\begin{equation}
A=4\pi r^2.
\end{equation}
A sphere is said to be {\em untrapped, marginal} or {\em trapped} respectively 
if $g^{-1}(dr)$ is spatial, null or temporal, where $g$ is the metric and 
$g^{-1}$ its inverse. If the space-time is time-orientable and $g^{-1}(dr)$ is 
future (respectively past) causal, then the sphere is said to be {\em future} 
(respectively {\em past}) trapped or marginal. A hypersurface foliated by 
marginal spheres is called a {\em trapping horizon} \cite{bhd,bhs}. 

The Kodama vector \cite{Kod} is 
\begin{equation}
k=g^{-1}({*}dr)
\end{equation}
where $*$ is the Hodge operator in the space normal to the spheres of symmetry, 
i.e.\ 
\begin{equation}
k\cdot dr=0,\qquad g(k,k)=-g^{-1}(dr,dr).
\end{equation}
This vector gives a preferred flow of time, coinciding with the static Killing 
vector of standard black holes such as Schwarzschild and Reissner-Nordstr\"om. 
Note that $k$ is temporal, null or spatial respectively on untrapped, marginal 
or trapped spheres. 

Both $k$ and the corresponding energy-momentum density 
\begin{equation}
j=-g^{-1}(T\cdot k)
\end{equation}
where $T$ denotes the energy-momentum-stress tensor, are covariantly conserved 
\cite{sph,1st}: 
\begin{eqnarray}
\nabla\cdot k&=&0\\
\nabla\cdot j&=&0
\end{eqnarray}
where $\nabla$ denotes the covariant derivative operator and the second 
property uses the Einstein equation. These Noether currents therefore admit 
Noether charges 
\begin{eqnarray}
V&=&-\int_\Sigma\hat{*}\cdot k\\
m&=&-\int_\Sigma\hat{*}\cdot j
\end{eqnarray}
where $\hat{*}$ denotes the volume form times unit normal of a spatial 
hypersurface with regular centre. The charges are found to be area volume 
\begin{equation}
V=\textstyle{\frac43}\pi r^3
\end{equation}
and the active gravitational mass $m$ \cite{MS}: 
\begin{equation}
1-2m/r=g^{-1}(dr,dr)
\end{equation}
where spatial metrics are positive definite and the Newtonian gravitational 
constant is unity. Evidently $r>2m$, $r=2m$ or $r<2m$ respectively on 
untrapped, marginal or trapped spheres. Various other properties illuminating 
the physical meaning of $m$ have been derived \cite{sph,1st,ine}. 

Surface gravity was defined as \cite{1st} 
\begin{equation}\label{kappa0}
\kappa=\textstyle{\frac12}{*}d{*}dr
\end{equation}
where $d$ is the exterior derivative in the normal space, i.e.\ ${*}d{*}d$ is a 
two-dimensional wave operator. It can be shown \cite{1st} to satisfy 
\begin{equation}
k\cdot(\nabla\wedge g(k))=\kappa dr
\end{equation}
and therefore
\begin{equation}
k\cdot(\nabla\wedge g(k))\cong\pm\kappa g(k)
\end{equation}
where $\cong$ henceforth denotes evaluation on a trapping horizon $r\cong2m$, 
similarly to the usual Killing identity. Then a trapping horizon is said to be 
{\em outer, degenerate} or {\em inner} respectively if $\kappa>0$, $\kappa=0$ 
or $\kappa<0$. Examples of all types are provided by Reissner-Nordstr\"om 
solutions. 

The Einstein equation implies
\begin{equation}\label{kappa1}
\kappa=\frac{m}{r^2}-4\pi r w
\end{equation}
where the work density is
\begin{equation}
w=-\textstyle{\frac12}\hbox{tr}\,T
\end{equation}
and the trace is in the normal space. In vacuo, $\kappa$ is $m/r^2$ and 
therefore reduces to the Newtonian surface gravity in the Newtonian limit, 
since $m$ reduces to the Newtonian mass \cite{sph,1st}. 

Another invariant of $T$ is the energy flux 
\begin{equation}
\psi=T\cdot g^{-1}(dr)+wdr.
\end{equation}
The Einstein equation implies
\begin{equation}\label{first}
dm=A\psi+wdV
\end{equation}
which was dubbed the unified first law \cite{1st}, as it encodes first laws of 
both thermodynamics and black-hole dynamics. Essentially, it expresses energy 
conservation, with the terms on the right-hand side being interpreted 
respectively as energy supply and work. 

The fields $(A,r,k,j,V,m,\kappa,w,\psi)$ have been introduced above in a 
manifestly invariant way. For calculations, it is often useful to use dual-null 
coordinates $x^{\pm}$, in terms of which any spherically symmetric metric can 
locally be written as 
\begin{equation}\label{metric}
ds^2=r^2d\Omega^2-2e^{2\varphi}dx^+dx^-
\end{equation}
where $d\Omega^2=d\theta^2+\sin^2\theta d\phi^2$ for spherical polar 
coordinates $(\theta,\phi)$, and $(r,\varphi)$ are functions of $(x^+,x^-)$. 
There is still the freedom to rescale functionally $x^{\pm}\to \tilde{x}^{\pm} 
(x^{\pm})$, under which $\varphi$ transforms by additive functions of $x^+$ and 
$x^-$. Then the following explicit expressions can be obtained: 
\begin{eqnarray}
&&k=e^{-2\varphi}(\partial_+r\partial_--\partial_-r\partial_+)\\
&&2m/r-1=2e^{-2\varphi}\partial_+r\partial_-r\\
&&\kappa=-e^{-2\varphi}\partial_+\partial_-r\label{kappa}\\ 
&&w=e^{-2\varphi}T_{+-}\\
&&\psi=-e^{-2\varphi}(T_{++}\partial_-rdx^++T_{--}\partial_+rdx^+).
\end{eqnarray}
This also shows how $\psi$ encodes radiative components of $T$, while $w$ 
encodes the Coulomb-like component, e.g.\ $w=E^2/8\pi$ for the electric field 
$E=q/r^2$ of a Reissner-Nordstr\"om black hole with charge $q$. 

The Einstein equations in these coordinates are 
\begin{eqnarray}
&&\partial_\pm\partial_\pm r-2\partial_\pm\varphi\partial_\pm r=-4\pi rT_{\pm\pm}
\label{focus}\\
&&r\partial_+\partial_-r+\partial_+r\partial_-r+\textstyle{\frac12}e^{2\varphi}
=4\pi r^2T_{+-}\\
&&r^2\partial_+\partial_-\varphi-\partial_+r\partial_-r-\textstyle{\frac12}e^{2\varphi}
=-4\pi r^2(T_{+-}+e^{2\varphi}p)\qquad
\end{eqnarray}
where $p=T_\theta^\theta=T_\phi^\phi$ is the transverse pressure. It follows 
that 
\begin{equation}
\partial_\pm m=4\pi r^2e^{-2\varphi}(T_{+-}\partial_\pm r-T_{\pm\pm}\partial_\mp r)
\end{equation}
which is a coordinate version of the unified first law (\ref{first}).

\section{Wormhole mouths}
A {\em wormhole mouth}, previously called horizon \cite{wh1}, is defined as a 
temporal outer trapping horizon. It is temporal in order to be two-way 
traversable, while the outer condition 
\begin{equation}
\kappa>0
\end{equation}
is proposed as the generalization of the minimality condition. Since 
$g^{-1}(dr,dr)=-2e^{-2\varphi}\partial_+r\partial_-r$, either $\partial_+r$ or 
$\partial_-r$ must vanish on the mouth, and $\partial_+r\cong0$ will be assumed 
henceforth. 

Introducing a generating vector $\xi$ of the marginal surfaces composing the 
mouth, its defining property is 
\begin{equation}\label{xi}
\xi\cdot d(\partial_+r)\cong0.
\end{equation}
Writing $\xi=\xi^+\partial_++\xi^-\partial_-$, one has 
\begin{equation}
\xi^+>0,\quad\xi^->0
\end{equation}
for future-pointing $\xi$. Then (\ref{xi}) expands as 
$\xi^+\partial_+\partial_+r+\xi^-\partial_-\partial_+r\cong0$, then 
(\ref{kappa}) shows that 
\begin{equation}\label{minimal}
\partial_+\partial_+r>0
\end{equation}
which expresses strict minimality of the sphere in the null hypersurface 
generated in the $\partial_+$ direction.

Hochberg \& Visser \cite{HV1,HV2} gave an alternative definition using a 
non-strict version of (\ref{minimal}) rather than $\kappa>0$. However, they 
specifically allowed spatial $\xi$, which does not give local two-way 
traversability and for which  (\ref{minimal}) can select maximal rather than 
minimal surfaces. For instance, consider any Robertson-Walker space-time with a 
bounce and a maximal surface in that time-symmetric hypersurface. There are two 
trapping horizons intersecting at the maximal surface, which if spatial satisfy 
(\ref{minimal}) there. 

Strict minimality has been assumed for simplicity. For a minimal surface in a 
null hypersurface, one has merely $\kappa\ge0$, which suffices for many 
purposes but also includes surfaces which are not minimal. The analysis to 
follow will be of a single wormhole mouth, though it should be stressed that 
two-way traversability requires two mouths with opposite senses, i.e.\ marginal 
in opposite null directions, in mutual causal contact. 

\section{Static wormholes}
Locally one can always introduce coordinates $(t,r_*)$ defined by 
$\sqrt2x^\pm=t\pm r_*$, where $r_*$ is a generalization of the Regge-Wheeler 
``tortoise'' coordinate \cite{RW}. Then the metric takes the form 
\begin{equation}
ds^2=r^2d\Omega^2+e^{2\varphi}(dr_*{}^2-dt^2).
\end{equation}
The metric in either $(x^+,x^-)$ or $(t,r_*)$ coordinates is manifestly regular 
if $(r,\varphi)$ are finite, assumed henceforth, and $r$ is non-zero. 

In a static case with static Killing vector $\partial_t$, so that $(r,\varphi)$ 
are independent of $t$, transforming further from $r_*$ to $r$ yields 
\begin{equation}
ds^2=r^2d\Omega^2+(1-2m/r)^{-1}dr^2-e^{2\varphi}dt^2
\end{equation}
which is essentially the Morris-Thorne form of the metric. They introduced new 
jargon which has been enthusiastically adopted by wormhole aficionados, namely 
``shape function'' for $2m$ and ``redshift function'' for $\varphi$. Indeed 
$\varphi$ is related to redshift, but better understood as a gravitational 
potential, reducing to the Newtonian potential in the Newtonian limit, if $t$ 
reduces to Newtonian time. This and the fact that $m$ is active gravitational 
mass are useful in physically interpreting such metrics. 

The Morris-Thorne metric is singular at the wormhole throat $r\cong2m$, so they 
used an embedding method to express minimality, as verified in the book of 
Visser \cite{Vis}. Actually, there is no need to use a fictitious embedding 
space, as minimality is an intrinsic property. While the Morris-Thorne paper is 
still in many ways an excellent read, this is one unfortunate aspect which 
continues to inspire confusion. In particular, it is not recommended to 
generalize by naively inserting time-dependent or angular-dependent factors 
into a metric which is singular precisely at the object of interest. 

In static cases, a wormhole mouth as defined above must be a double trapping 
horizon, $\partial_+r\cong\partial_-r\cong0$, since 
$\partial_\pm=\sqrt2(\partial_t\pm\partial_*)$ and $\partial_tr=0$. Then 
\begin{equation}
\partial_*r\cong0
\end{equation}
so the surface is extremal in the static hypersurface. Since 
$\partial_t\partial_tr=0$, one finds 
$\partial_*\partial_*r=2\partial_+\partial_+r=2\partial_-\partial_-r=-2\partial_+\partial_-r$, 
so the minimality condition (\ref{minimal}) implies that the surface is 
strictly minimal in the static hypersurface, 
\begin{equation}
\partial_*\partial_*r>0.
\end{equation}
Thus the proposed definition of wormhole mouth recovers the Morris-Thorne 
definition appropriately. One may equivalently use proper radius $\int 
e^\varphi dr_*$ instead of $r_*$ \cite{Vis}. 

A calculation shows that 
\begin{equation}
2\kappa=\frac{m}{r^2}-\frac{\partial_rm}{r}+\left(1-\frac{2m}{r}\right)\partial_r\varphi.
\end{equation}
In particular, 
\begin{equation}
2\kappa\cong(m-r\partial_rm)/r^2
\end{equation}
so that $\kappa>0$ reduces to the ``flaring-out'' condition of Morris \& 
Thorne, their equation (54). Thus their embedding method correctly expresses 
strict minimality in static cases. 

Another calculation shows that 
\begin{equation}
\kappa\cong2\pi r(\tau-\rho)
\end{equation}
where $\rho=-T_t^t$ is the energy density and $\tau=-T_r^r$ the radial tension. 
This confirms that the weak energy condition must be violated \cite{MT}, 
\begin{equation}
\tau>\rho.
\end{equation}
Moreover, it provides a simple measure of the violation, $\tau-\rho$ equalling  
surface gravity over circumference \cite{wh1}. 

One might object to referring to $\kappa$ as surface gravity in the case of a 
static wormhole; it is rather an invariant measure of the radial curvature or 
``flaring-out''. Quoting tetrad Riemann curvature components in equations (8) 
of Morris \& Thorne: 
\begin{equation}
R^{\hat r}{}_{\hat\theta\hat r\hat\theta}=R^{\hat r}{}_{\hat\phi\hat r\hat\phi}
\cong-2\kappa/r.
\end{equation}
While Riemann curvature, like Gaussian curvature, has units of inverse length 
squared, $\kappa$, like principal curvature, has units of inverse length, as 
appropriate for measuring radial curvature. Actually, anywhere in a static 
space-time, the same expressions show that 
\begin{equation}
R^{\hat r}{}_{\hat\theta\hat r\hat\theta}+R^{\hat t}{}_{\hat\theta\hat t\hat\theta}
=R^{\hat r}{}_{\hat\phi\hat r\hat\phi}+R^{\hat t}{}_{\hat\phi\hat t\hat\phi}
=-2\kappa/r
\end{equation}
which indicates how $\kappa$ generally also includes temporal curvature.

\section{Laws of wormhole dynamics}

This section derives some basic laws of wormhole dynamics which were stated 
previously \cite{wh1}. Applications to dynamical processes involving wormholes 
have been made concrete in various examples, so will be only briefly mentioned 
and cited. 

{\em Negative energy density:} the null energy condition is necessarily 
violated on a wormhole mouth. Proof: use the focussing equations (\ref{focus}) 
and the minimality property (\ref{minimal}): $T_{++}<0$. 

This confirms that ``exotic'' matter is required even in dynamic cases, which 
might also be called phantom or ghost matter, generalizing terminology from 
cosmology or quantum field theory respectively. Claims that this need not be so 
in Einstein gravity either do not involve traversable wormholes as defined 
here, or evaluate energy density somewhere other than a mouth, e.g.\ at a 
centre of symmetry described as a ``throat'', or involve calculational errors. 
In dynamic cases, there are actually two independent constraints on energy 
density \cite{IH}. 

Since a black hole may be locally defined by a future outer trapping horizon 
\cite{bhd,1st}, this allows interconversion of black holes and traversable 
wormholes. A future outer trapping horizon characterizes a black hole if 
achronal, equivalently $T_{++}\ge0$, and a wormhole if temporal, equivalently 
$T_{++}<0$. Thus a traversable wormhole can collapse to a black hole if its 
negative-energy source fails, or if enough positive-energy matter or radiation 
is pumped in \cite{wh1,wh2,wh3,wh4,wh5,wh6,wh8}. Conversely, a black hole can 
be converted into a traversable wormhole by beaming in enough negative-energy 
radiation \cite{wh1,wh2,wh5,wh6,wh7,wh8,wh9,wh10}. Wormhole construction from 
disjoint regions of flat space-time has recently been demonstrated by Maeda 
\cite{Mae}, albeit with a singularity at the topology change. 

{\em Zeroth law:} $\kappa$ is constant on a static wormhole throat. Proof: 
obvious. 

{\em Second law:} future, past or static wormhole mouths respectively have 
decreasing, increasing or constant area. Proof: the expansion of the mouth is 
\begin{equation}
A'/A=2r'/r
\end{equation}
where the prime henceforth denotes $\xi\cdot d$. One can expand $\xi\cdot 
dr=\xi^+\partial_+r+\xi^-\partial_-r\cong\xi^-\partial_-r$, while $\partial_-r$ 
is negative, positive or zero respectively for future, past or static wormhole 
mouths. 

This is like the second law of black-hole dynamics \cite{bhd}, but with 
reversed sign, reflecting the causal character of the mouth, or equivalently, 
the reversed null energy condition. It follows that a static wormhole is 
enlarged or reduced, respectively, by opening then closing a region of past or 
future trapped surfaces. To enlarge, this can be done by beaming in negative 
energy, balanced by subsequent positive energy, while the opposite order would 
reduce the area \cite{wh1,wh3,wh4,wh6,wh7,wh8,wh9,wh10}. For some matter 
models, this can be done in an apparently stable way 
\cite{wh2,wh3,wh5,wh6,wh7,wh8,wh9,wh10}, while others are unstable, leading 
either to collapse to a black hole as above, or to inflationary expansion 
\cite{wh4,wh8,GGS1,GGS2,DHNS}. 

{\em First law:}
\begin{equation}
m'\cong\frac{\kappa A'}{8\pi}+wV'.
\end{equation}
Proof: one can calculate this by a few steps as the projection of the unified 
first law (\ref{first}) along $\xi$, but the easiest way is to multiply the 
expression (\ref{kappa1}) for $\kappa$ by $A'$, then use the fact that 
$r\cong2m$ and $r'\cong2m'$. 

This has the same form as the first law of black-hole dynamics \cite{1st}. 
These three laws therefore suggest a genuine connection with thermodynamics. 
Indeed, it has recently been shown that any future outer trapping horizon has a 
local Hawking temperature $\kappa/2\pi$ \cite{bht}, which therefore applies to 
future wormhole mouths. Thus there is a new field of wormhole thermodynamics 
waiting to be explored. 

\medskip 
Research supported by the National Natural Science Foundation of China under 
grants 10375081, 10473007 and 10771140, by Shanghai Municipal Education 
Commission under grant 06DZ111, and by Shanghai Normal University under grant 
PL609.

\end{document}